\documentclass[aps, reprint, superscriptaddress]{revtex4-1}
\usepackage{amsmath}
\usepackage{amssymb}
\usepackage{bm}
\usepackage{graphicx}
\usepackage{hyperref}

\newcommand{\mvec}[1]{\ensuremath{\mathbf{#1}}}
\newcommand{\mvect}[2]{\ensuremath{\mathbf{#1}_\mathrm{#2}}}
\newcommand{\Heff}{\ensuremath{\mvec{H}_\mathrm{eff}}}
\newcommand{\Hd}{H_{\!D}}
\newcommand{\Hdhat}{\hat{H}_{\!D}}

\newcommand{\up}{u_{\mathrm{pos}}}
\newcommand{\un}{u_{\mathrm{neg}}}

\newcommand{\ex}{\ensuremath{\mvect{e}{x}}}
\newcommand{\ey}{\ensuremath{\mvect{e}{y}}}
\newcommand{\ez}{\ensuremath{\mvect{e}{z}}}

\newcommand{\mean}[1]{\langle {#1} \rangle}

\begin{document}
\title{Magnonic analog of relativistic Zitterbewegung in an antiferromagnetic spin chain}

\date{\today}

\author{Weiwei Wang}
\email{wangweiwei1@nbu.edu.cn}
\affiliation{Faculty of Science, Ningbo University, Ningbo 315211, China}
\author{Chenjie Gu}
\affiliation{Faculty of Science, Ningbo University, Ningbo 315211, China}
\author{Yan Zhou}
\affiliation{School of Science and Engineering, Chinese University of Hong Kong (Shenzhen), China}
\author{Hans Fangohr}
\affiliation{Engineering and the Environment, University of Southampton,  SO17 1BJ, Southampton, United Kingdom}

\begin{abstract}
We theoretically investigate the spin wave (magnon) excitations in a classical antiferromagnetic spin chain
with easy-axis anisotropy. We obtain a Dirac-like equation by linearizing the Landau-Lifshitz-Gilbert equation
in this antiferromagnetic system, in contrast to the ferromagnetic system in which a Schr\"{o}dinger-type
equation is derived. The Hamiltonian operator in the Dirac-like equation is a pseudo-Hermitian.
We compute and demonstrate the relativistic Zitterbewegung (trembling motion) in the antiferromagnetic spin chain
by measuring the expectation values of the wave packet position.
\end{abstract}

\maketitle

\section{Introduction}
As the fundamental equation for relativistic quantum mechanics, the Dirac equation~\cite{Thaller1992}
provides a natural description of a spin-1/2 massive particle~\cite{Gerritsma2010} and
manages to reproduce the spectrum of the hydrogen atom accurately~\cite{Lamata2007}.
Meanwhile, the Dirac equation also predicts some astonishing effects,
such as the \textit{Zitterbewegung} (ZB)
and Klein's paradox. The ZB is an unexpected trembling motion of a free relativistic
quantum particle~\cite{Thaller1992, Gerritsma2010, Lamata2007, Longhi2010}, originated from
the interference of positive and negative energy states~\cite{Thaller2005, Gerritsma2010}.
However, it is tough to observe the ZB directly for relativistic electrons because the amplitude
of ZB oscillations is tiny (of the order of the Compton wavelength)~\cite{Huang1952, Vaishnav2008, Longhi2010}
and the oscillation frequency is extremely high ($\approx 10^{21}$ Hz)~\cite{Dreisow2010}.
Therefore, much work has been devoted to simulating ZB in controllable classical and quantum systems.
For example, photonic analogs of the ZB have been proposed~\cite{Zhang2008a, Longhi2010} and
successfully demonstrated in the experiment using photonic lattices~\cite{Dreisow2010}.
A quantum simulation employing trapped ions~\cite{Lamata2007} is performed to illustrate the ZB as well~\cite{Gerritsma2010}.
Moreover, various systems such as ultracold neutral atoms~\cite{Vaishnav2008}, graphene~\cite{Cserti2006}, metamaterials~\cite{Ahrens2015} and Bose-Einstein condensate (BEC)~\cite{Leblanc2013} were
proposed as candidate systems to observe the ZB. In this paper, we show that the classical
analog of relativistic ZB could occur in antiferromagnetic systems.

The emergence of antiferromagnetic spintronics~\cite{Gomonay2014, Jungwirth2016} have drawn considerable
attentions recently, such as the reports of spin Seebeck effect~\cite{Wu2016, Rezende2016a}, spin pumping
and spin-transfer torques~\cite{Cheng2014} in antiferromagnets, the antiferromagnetic domain wall dynamics
driven by magnons (spin waves)~\cite{Tveten2014} as well as spin-orbit torques~\cite{Gomonay2016, Shiino2016},
and the electrical switching of an antiferromagnet~\cite{Wadley2016}. Comparing with the spin waves
in ferromagnetic materials, spin waves in antiferromagnetic materials provide more degrees of freedom
for information encoding and processing~\cite{Cheng2016}, which can be seen from the fact that the
Schr\"{o}dinger equation is obtained in the ferromagnetic system~\cite{Yan2011} while a two-component
Klein-Gordon equation is derived for the antiferromagnetic system~\cite{Cheng2016}.
A very popular method to describe the antiferromagnetic system is to use $\mvec{M}$ (uncompensated magnetization)
and the N\'{e}el order parameter $\mvec{L}$ (staggered magnetization)~\cite{Ivanov1995, Gomonay2016, Shiino2016, Cheng2016}. Interestingly, an intrinsic magnetization emerges if the order parameter varies spatially~\cite{Tveten2016}.
In this work, we will make use of the classical spin model and linearize the Landau-Lifshitz-Gilbert (LLG)
equation directly without bothering $\mvec{M}$ and $\mvec{L}$. We will show that a Dirac-like equation
is obtained and based on which we demonstrate the relativistic effects such as the ZB,
analytically and numerically.

\section{Model}
\begin{figure}[tbhp]
\begin{center}
\includegraphics[width=0.48\textwidth]{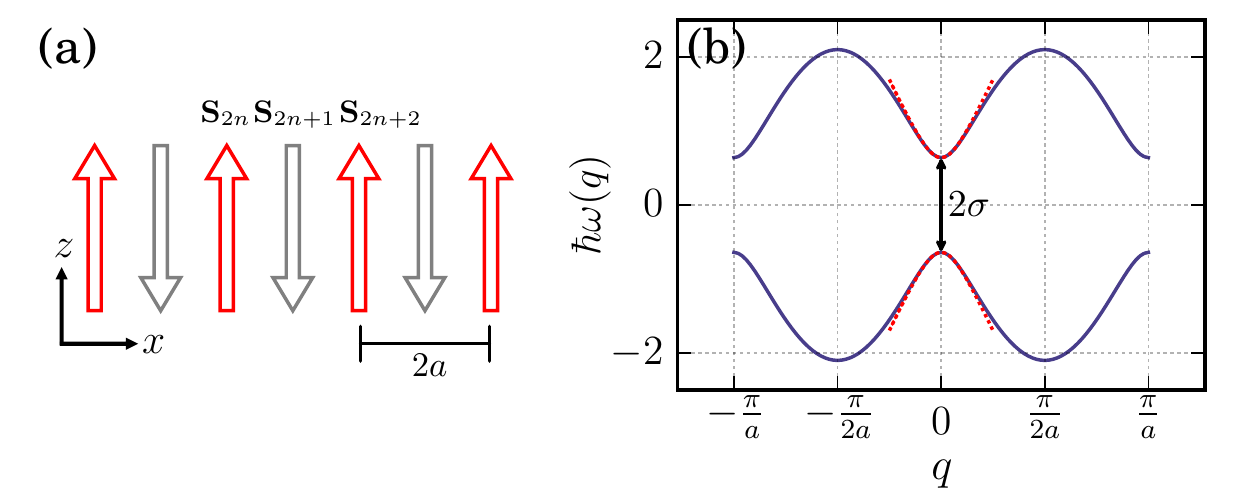}
\caption{(a) Illustration of a two-sublattice antiferromagnetic spin chain with lattice constant $2a$.
The ground state is the spins antiparallel with each other along the $z$-axis.
(b) The dispersion relations of the antiferromagnetic lattice, consisting of two branches separated by $2\sigma$.
The dotted curves show dispersion relations in the long-wavelength limit,
and in the plotting $J=1.0$, $D=0.05J$ and $H=0$ are used.}
\label{fig_afm}
\end{center}
\end{figure}

We consider a classical Heisenberg spin chain along the $x$-direction with energy contributions from
antiferromagnetic exchange interaction, an uniaxial anisotropy and an external Zeeman field.
The Hamiltonian of the one-dimensional system is given by~\cite{Haldane1983, Ivanov1995, Kampfrath2011, Rezende2016a}
\begin{equation}\label{eq_ham}
\mathcal{H}= J \sum_{n} \mvec{S}_n\cdot \mvec{S}_{n+1} - D \sum_{n}  (\mvec{e}_z \cdot \mvec{S}_n)^2 - H \sum_{n} \mvec{e}_z \cdot \mvec{S}_n
\end{equation}
where $J (>0)$ is the exchange constant, $D$ denotes the uniaxial anisotropy strength and $H$ represents the Zeeman field.
The spin $\mvec{S}_{n}$ is treated as a classical vector with length $S$ and the associated magnetic moment
is $\mvec{\mu}_s=\hbar \gamma S$. 
This model is valid for easy-axis antiferromagnets with sufficiently large spin~\cite{Haldane1983, Ivanov1995}.
The spin dynamics at lattice site $n$ is governed by the LLG equation,
\begin{equation}\label{eq_llg}
\frac{\partial \mvec{S}_n}{\partial t} = - \gamma \mvec{S}_i \times \Heff + \frac{\alpha}{S} \mvec{S}_n \times  \frac{\partial \mvec{S}_n}{\partial t}
\end{equation}
where $\alpha$ is the Gilbert damping and $\Heff$ is the effective
field that is computed as $\Heff = -(1/\hbar \gamma) ({\partial \mathcal{H}}/{\partial \mvec{S}_n})$.
In the ground state, the spin $\mvec{S}_{n}$ is antiparallel to its neighbours $\mvec{S}_{n\pm1}$,
i.e., $\mvec{S}_{n}=(-1)^n S\mvec{e}_z$, as shown in Fig.~\ref{fig_afm}(a).

\subsection{Linearizing the LLG equation}
We assume that the spin wave (magnon) excitation can be described by a small fluctuation around the ground state:
$\mvec{u}_n=u_x(n) \ex + u_y(n) \ey$ and $\mvec{v}_n=v_x(n) \ex + v_y(n) \ey$ where
$|\mvec{u}|\ll 1$ and $|\mvec{v}|\ll 1$. So we have $\mvec{S}_{2n}=S(\mvec{u}_n + \sqrt{1-u^2}\ez)$
and $\mvec{S}_{2n+1}=S(\mvec{v}_n - \sqrt{1-v^2}\ez)$.
By introducing the complex variables $\psi^a_n= u^x_n+i u^y_n$ and $\psi^b_n= v^x_n+i v^y_n$,
linearizing the LLG equation with $\alpha=0$ we obtain
\begin{align}\label{eq_llg2}
i \hbar \frac{d \psi^a_n}{d t} &= -J_s (\psi^b_{n-1}+\psi^b_n)-(2J_d+H) \psi^a_n, \nonumber\\
i \hbar \frac{d \psi^b_n}{d t} &= J_s (\psi^a_{n+1}+\psi^a_n)+(2J_d-H) \psi^b_n.
\end{align}
where $J_s=JS$ and $J_d=(J+D)S$. Using the plane wave ansatz $\psi^{a(b)}_n \sim \exp(i q n2a - i \omega t)$,
the dispersion relations  of spin waves can be found as
\begin{equation}\label{eq_dis_q}
\hbar \omega(q)= -H \pm S\sqrt{4D(D+2J)+4J^2 \sin^2 (qa)}.
\end{equation}
Fig.~\ref{fig_afm}(b) plots the dispersion relations of the two-sublattice antiferromagnetic spin chain
with $D=0.05J$ and $H=0$. The two branches are separated by $2\sigma$ where $\sigma=S\sqrt{4D(D+2J)}$.

In the continuum approximation, we introduce a dimensionless coordinate $\xi$ to approach $n$, i.e.,
\begin{equation}
\xi \leftrightarrow n=x/(2a).
\end{equation}
For the case that both $\psi^a$ and $\psi^b$ vary slowly with $n$, we have
$\psi^a_{n+1} \approx \psi^a_n+ \partial_\xi \psi^a_n + (1/2)\partial_{\xi \xi} \psi^a_n$ and
$\psi^b_{n-1} \approx \psi^b_n- \partial_\xi \psi^b_n + (1/2)\partial_{\xi \xi} \psi^b_n$.
It is worth noting that we have included the second order derivatives because without them
the correct dispersion relations can not be reached. Moreover, the second order derivatives
emerge naturally by using the central difference scheme if one introduces new variables
$\psi^c_n=\psi^a+i \psi^b$ and $\psi^d_n=\psi^a-i \psi^b$.
With the two-component Dirac spinor $\psi(\xi, t)=(\psi_1, \psi_2)^T=(\psi^a, \psi^b)^T$,
Eq.(\ref{eq_llg2}) can be cast into a Dirac-like equation:
\begin{equation} \label{eq_dirac}
i \hbar \frac{\partial \psi}{\partial t}= (\Hd+V) \psi
\end{equation}
where $V=-H$ plays the role of potential. The corresponding Dirac Hamiltonian $\Hd$ is
\begin{equation} \label{eq_Hd}
\Hd = J_s \sigma_x \partial_\xi  -i J_s \sigma_y [2
+ (1/2) \partial_{\xi \xi}] -  2 J_d \sigma_z
\end{equation}
where 
\begin{equation}
\sigma_x = \left[ \begin{array}{cc} 0 & 1 \\ 1 & 0 \end{array} \right],
\quad
\sigma_y = \left[ \begin{array}{cc} 0 & -i \\ i & 0 \end{array} \right],\quad
\sigma_z = \left[ \begin{array}{cc} 1 & 0 \\ 0 & -1 \end{array} \right]
\end{equation}
are Pauli matrices. Interestingly, $\Hd$ is not a 
Hermitian operator, i.e., $\Hd\neq \Hd^\dagger$ where the symbol $\dagger$ 
denotes the usual Dirac Hermitian conjugation (transpose and complex conjugate).
Note that $\sigma_i=\sigma_i^\dagger$ ($i=x,y,z$) and 
$\partial_\xi^\dagger=-\partial_\xi$ so we have
$\Hd^\dagger = -J_s \sigma_x \partial_\xi  +i J_s \sigma_y [2 + (1/2) \partial_{\xi \xi}] -  2 J_d \sigma_z $ 
and thus $\Hd+\Hd^\dagger=-4J_d \sigma_z$. 

In the long-wavelength limit, the dispersion relation (\ref{eq_dis_q}) reduces to
\begin{equation} \label{eq_dispersion}
\hbar \omega(k)\approx -H \pm \sqrt{\sigma^2+ J_s^2 k^2},
\end{equation}
where $k=2aq$. The dispersion [Eq.~(\ref{eq_dispersion})] agrees with the one obtained 
in the approach of Holstein-Primakoff~\cite{Rezende2016, Rezende2016a}.  
In the absence of external field, we have $\hbar \omega(k)=\pm \epsilon(k)$ with
$\epsilon(k)=\sqrt{\sigma^2+ J_s^2 k^2}$. In Fig.~\ref{fig_afm}(b) the dotted lines show
the dispersion relations in the long-wavelength limit, which mimic the typical energy-momentum
dispersion relations with positive and negative energy states of a relativistic free massive particle.

It is of interest to connect Eq.~(\ref{eq_dirac}) with the Klein-Gordon equation. For example, 
in the absence of external field Eq.~(\ref{eq_dirac}) can be rewritten as
\begin{equation} \label{eq_kg}
(J_s^2 \partial_\xi^2-\hbar^2 \partial_t^2)\psi=\sigma^2 \psi
\end{equation}
where the high-order derivative is neglected. In terms of small deviations $\mvec{n}_\perp=\{n_x, n_y\}$ 
of the staggered field defined by $\mvec{n}=(\mvec{S}_{2n}-\mvec{S}_{2n+1})/(2S)$, one has
$(J_s^2 \partial_\xi^2-\hbar^2 \partial_t^2)\psi_+=\sigma^2 \psi_+$
where $\psi_+=\psi^a_n-\psi^b_n=n_x+i n_y$.

\subsection{Eigenvectors and plane-wave solutions}
In the momentum space, the Dirac spinor $\hat{\psi}$ can be obtained by applying the 
Fourier transformation to each component of Dirac spinor $\psi$~\cite{Thaller2005},
\begin{equation}
\hat{\psi}_j(k) = (\mathcal{F} \psi_j) (k) = \frac{1}{\sqrt{2\pi}} \int e^{-ik\xi} \psi_j(\xi) d\xi
\end{equation} 
where $\mathcal{F}$ denotes the Fourier transformation operator. Therefore,
$(\mathcal{F} \Hd \psi)(k) = \Hdhat (\mathcal{F} \psi) (k)$
and $\Hdhat(k)=\mathcal{F} \Hd \mathcal{F}^{-1}$ where
\begin{equation}
\Hdhat(k) = ikJ_s \sigma_x - iJ_s \sigma_y(2-k^2/2)-2J_d\sigma_z.
\end{equation} 
Apparently, $\Hdhat$ can be rewritten in the matrix form 
\begin{equation}
\Hdhat = \left[ \begin{array}{cc} -d_z &  i d_x-d_y \\ i d_x+ d_y& d_z \end{array} \right]
\end{equation}
where $d_x=J_s k$, $d_y=J_s(2-k^2/2)$ and $d_z=2 J_d$. Clearly, $\Hdhat$ is not a Hermitian matrix; 
its normalized eigenvectors are given by 
\begin{equation}
u_+(k, \epsilon)=\frac{1}{\sqrt{2 d_z (d_z+\epsilon)}}\left[ \begin{array}{c} id_x-d_y \\ d_z+\epsilon 
 \end{array} \right]
\end{equation}
and $u_-(k, \epsilon)=u_+(k, -\epsilon)$.
The  eigenvalues $\lambda=\pm \epsilon(k)$ with $\epsilon=\sqrt{d_z^2-d_y^2-d_x^2}$ 
correspond to the positive and negative energy states. 
Note that $u_\pm$ are not orthogonal, i.e., $\mean{u_+, u_-} \neq 0$ where 
the Hermitian inner product of two vectors $\psi=(\psi_1, \psi_2)^T$ and $\phi=(\phi_1, \phi_2)^T$
is defined by
\begin{equation}\label{eq_inner1}
\mean{\psi, \phi}= \psi^* \cdot \phi = \psi_1^*\phi_1+\psi_2^*\phi_2.
\end{equation}

In the long-wavelength limit,
$\epsilon \approx \sqrt{\sigma^2+ J_s^2 k^2}$ where $\sigma=S\sqrt{4D(D+2J)}$.
The normalized eigenvectors thus can be approximately given by
\begin{equation}\label{eq_evector}
u_+(k, \epsilon )=\left[ \begin{array}{c} d_1(k, \epsilon) \\ d_2(k, \epsilon)
 \end{array} \right], \quad
 u_-(k, \epsilon)=u_+(k, -\epsilon)
\end{equation}
where $d_1(k, \epsilon)=(-1/2+ik/4) (2 - \epsilon/J_d)^{\frac{1}{2}} $ and
$d_2(k, \epsilon)= (J/J_d)(2 - \epsilon/J_d)^{-\frac{1}{2}}$.
Based on the eigenvectors, it is straightforward to construct the plane-wave solutions.
Note that $\Hd e^{ik\xi}= \Hdhat(k) e^{ik\xi}$, so the stationary plane waves
$\up(k, \xi)=u_+(k) e^{ik\xi}$ and $\un(k, \xi)=u_-(k) e^{ik\xi}$ are the eigenfunctions of $\Hd$. 
Therefore, the plane-wave solutions of Eq.~(\ref{eq_dirac}) are
\begin{align}
u_p(k, \xi, t)= u_+ e^{i k \xi - i \epsilon(k) t/\hbar} \nonumber\\
u_n(k, \xi, t)= u_- e^{i k \xi + i \epsilon(k) t/\hbar}
\end{align}
The solution $u_p$ ($u_n$) is considered as a plane wave with positive (negative) energy~\cite{Thaller2005} since its eigenvalue is positive (negative).
As illustrated in Fig.\ref{fig_afm}(b), waves with positive (negative) energy have a 
positive (negative) group velocity if they have a positive wave vector.

Knowing the plane wave solutions $u_p$ and $u_n$, the general solution of the Dirac-type equation
(\ref{eq_dirac}) can be constructed,
\begin{equation}\label{eq_sol}
\psi(\xi, t)=\frac{1}{\sqrt{2\pi}}\int [\hat{f} u_+ e^{-i \epsilon t /\hbar} +  \hat{g} u_- e^{i \epsilon t /\hbar}] e^{i k \xi} dk
\end{equation}
where $\hat{f}=\hat{f}(k)$ and $\hat{g}=\hat{g}(k)$ are parameters to be determined 
by the initial condition $\psi(\xi, 0)=\psi_0(\xi)$.
It is clear that $\hat{f}(k) u_+(k)$ and $\hat{g}(k) u_-(k)$ denote the static wavefunctions that
have positive and negative energies, respectively. Gaussian Dirac spinors with pure positive or
negative energy can be built directly by setting $\hat{f}(k)$ and $\hat{g}(k)$ to be a Gaussian function.
With the helper function $\hat{\psi}_0(k)=(1/\sqrt{2\pi})\int \psi_0(\xi) e^{- i \xi k} d\xi$, we find that
$\hat{f}(k)=[\mean{u_-,\hat{\psi}_0} C - \mean{u_+,\hat{\psi}_0}]/(C^*C-1)$
and $g(k)=[\mean{u_+,\hat{\psi}_0} C^* - \mean{u_-,\hat{\psi}_0}]/(C^*C-1)$
where $C=\mean{u_+,u_-}$ and in the long-wavelength limit $C\approx J_s/J_d$.

\subsection{Operators and Expectation Values}
The operator $\Hdhat$ is not a Hermitian, it is a pseudo-Hermitian~\cite{Mostafazadeh2010}. 
A linear operator $B$ is said to be pseudo-Hermitian if there exists a Hermitian operator $\eta$
satisfies that $B^\dagger= \eta B \eta^{-1}$.
It is easy to check that $\Hdhat^\dagger =\sigma_z \Hdhat \sigma_z^{-1}$, therefore, $\Hdhat$ is 
$\Sigma_3$-pseudo-Hermitian.
However, the usual Dirac Hermitian conjugation (transpose and complex conjugate) actually 
is associated with the inner product [Eq.~(\ref{eq_inner1})], for instance,
the adjoint of an operator $\Omega^\dagger$ is defined through~\cite{Das2011}
\begin{equation}\label{eq_adjoint1}
\mean{\psi, \Omega \phi} = \mean{\Omega^\dagger \psi, \phi}.
\end{equation}
This suggest that the hermiticity and unitarity of operators are related 
to the choice of inner product~\cite{Pauli1943, Feshbach1958, Mostafazadeh2010, Das2011}.
For a pseudo-Hermitian operator, a natural choice of the inner product is
\begin{equation}\label{eq_inner2}
\mean{\psi, \phi}_\eta=\mean{\psi, \eta\phi} = \psi^* \cdot \eta \phi.
\end{equation}
For instance, if we choose $\eta=\sigma_z$ for $\Hdhat$ we arrive at
\begin{align}
\mean{\psi, \Hdhat \phi}_\eta&=\mean{\psi, \Hdhat' \phi}
= \mean{\Hdhat' \psi,  \phi} =\mean{\Hdhat \psi,  \phi}_\eta,
\end{align}
where we have used $\Hdhat'=\sigma_z \Hdhat$ is Hermitian. 
Therefore, the inner product [Eq.~(\ref{eq_inner2})] renders $\Hdhat$ Hermitian,
i.e., $\Hdhat^\star=\Hdhat$ where the symbol $^\star$ represents the adjoint of an operator 
and $\Omega^{\star}$ is defined by
\begin{equation}\label{eq_adjoint2}
\mean{\psi, \Omega \phi}_\eta = \mean{\Omega^\star \psi, \phi}_\eta.
\end{equation}
The two definitions [Eq.(\ref{eq_adjoint1})] and [Eq.(\ref{eq_adjoint2})] are connected to each other 
through~\cite{Feshbach1958,Das2011}
\begin{equation}
\Omega^{\star} = \eta^{-1} \Omega^\dagger \eta.
\end{equation}
Eigenfunctions of hermitian operators should be orthogonal, indeed, $u_\pm$ are orthogonal 
with respect to the inner product $\mean{\psi, \phi}_{\sigma_z}$, i.e., 
$\mean{u_+, u_-}_{\sigma_z}=0$.

\begin{table}
\caption{\label{tab1} Conservation table for different energy states.}
\begin{ruledtabular}
\begin{tabular}{ccc}
 Quantity & $\int \mean{\psi, \psi} d\xi$ & $\int \mean{\psi, \psi}_{\sigma_z} d\xi$\\
\hline
Pure positive energy & $\surd$ & $\surd$ \\
Pure negative energy & $\surd$ & $\surd$ \\
Mixed energies & $\times$ & $\surd$  \\
\end{tabular}
\end{ruledtabular}
\end{table}

In line with the inner product [Eq.~(\ref{eq_inner2})], the expectation value of an operator $\Omega$ thus is defined by~\cite{Pauli1943, Feshbach1958}
\begin{equation}\label{eq_expection}
\mean{\Omega}_{\sigma_z} =\int \psi^* \sigma_z \Omega \psi d \xi.
\end{equation}
This definition is in agreement with the hermiticity condition $\Omega^{\star} = \Omega$. 
Also, Eq.(\ref{eq_expection}) immediately implies the conservation of normalization $N=\int \mean{\psi, \psi}_{\sigma_z} d\xi=\int (\psi_1^* \psi_1 - \psi_2^* \psi_2) d\xi$. 
In general the quantity $N_+=\int \mean{\psi, \psi} d\xi =\int (\psi_1^* \psi_1 + \psi_2^* \psi_2) d\xi$ is not conserved.
However, for plane waves and wave packets with pure positive or negative energy, $N_+$ is conserved as well, 
as shown in Table~{\ref{tab1}}. With respect to the inner product [Eq.~(\ref{eq_inner2})], the transformation
\begin{equation}
\hat{S} = e^{-i \Hdhat t/\hbar}
\end{equation}
is unitary~\cite{Feshbach1958}. The time dependence of a wave function $\psi(t)$ reads
\begin{equation}
\psi(t)= e^{-i \Hdhat t/\hbar} \psi(0),
\end{equation}
and the time dependence of the expectation value of the  operator $\Omega$ is given by
\begin{equation}
\frac{d}{dt} \mean{\Omega}_{\sigma_z} =\frac{i}{\hbar} \mean{\Hd \Omega-\Omega \Hd}_{\sigma_z}.
\end{equation}

\section{Results}

We perform numerical simulations by solving Eqs.~(\ref{eq_ham}-\ref{eq_llg}) directly~\cite{Fidimag}.
We have chosen $J = \hbar  = \gamma = S = a = 1$ as the simulation parameters~\cite{Wang2015b},
and the Gilbert damping $\alpha=10^{-5}$ is used. The external field $H=0$ and $D=0.01J$.
We initialize the system with a Gaussian wave packet located at $\xi=0$, i.e., $\psi(\xi)=G_0(\xi) (1,0)^T$
where $G_0(\xi) = A \sqrt{\pi/L} e^{-\xi^2/(4L)} e^{i k_0 \xi}$ with $A=0.1$, $L=4000$ and $k_0=0.05$,
as shown in Fig.\ref{fig_zba}(a). The packet starts to oscillate immediately and splits into
two parts (packets) eventually. One moves towards left while the other moves right.
The packet that moves left (right) has negative (positive) energy.
Interestingly, the amplitudes of two parts are much larger than the original one,
which is because the system is dominated by antiferromagnetic exchange interaction and part of the
exchange energy has transferred into the anisotropy energy. It is worth mentioning that the wave packets
have a long decay (spreading) time, especially for the $D=0$ case, in contrast to
the packets in ferromagnetic system~\cite{Wang2016}. The detailed splitting process can be found
in the video [I.gif] of the Supplemental Material~\cite{Supp}.

The normalization factor $N=\int (\psi_1^* \psi_1 - \psi_2^* \psi_2) d\xi$ as well as the two individual components 
$\int \psi_1^* \psi_1 d\xi$ and $\int \psi_2^* \psi_2 d\xi$ are plotted in Fig.\ref{fig_zba}(b). It is found 
that the two components show a damped oscillation while the normalization factor $N$ is almost a constant.
Moreover, the normalization factor $N$ is much smaller than its components. 

\subsection{Zitterbewegung}
\begin{figure}[tbhp]
\begin{center}
\includegraphics[width=0.48\textwidth]{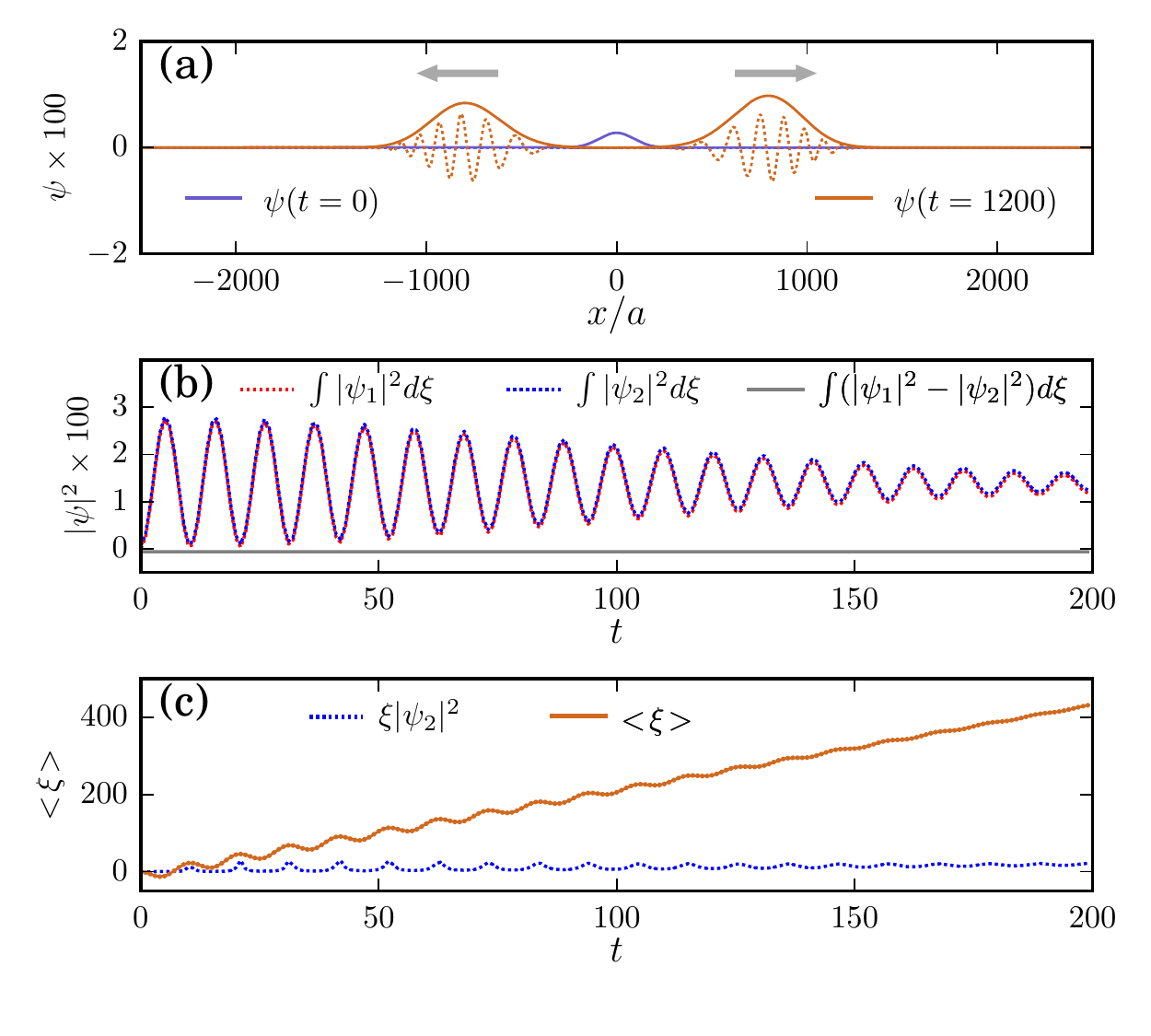}
\caption{(a) A Gaussian wave packet (blue line for $t=0$) splits into two (chocolate line for $t=1200$),
one moves towards left while the other moves towards right. The dashed lines plot $\mathrm{Re}(\psi_1)$,
i.e., the real part of the first component of $\psi$.
(b) The normalization factor $N=\int (\psi_1^* \psi_1 - \psi_2^* \psi_2) d\xi$ is almost a constant during 
the simulation.
(c) The normalized average value of $\mean{\xi}(t)$ as a function of time, where
 $\xi|\psi_1|^2$ and  $\xi|\psi_2|^2$ are the average values of its two components.}
\label{fig_zba}
\end{center}
\end{figure}

The ZB arises from the expectation value of the standard position operator, i.e.,
\begin{equation}
\mean{\xi} \equiv \mean{\xi}_{\sigma_z}= \int  \xi (|\psi_1|^2-|\psi_2|^2) d\xi.
\end{equation}
The usual method to deal with ZB is to derive the time-dependent equations in the Heisenberg 
picture~\cite{Thaller1992, Longhi2010}.
Here, instead of deriving the position operator in the Heisenberg picture,
we work directly in the Schr\"{o}dinger picture~\cite{Longhi2010}.
Starting from the wavefunctions [Eq.(\ref{eq_sol})] in momentum space, i.e., 
$\psi(\xi, t)=(1/\sqrt{2\pi})\int \hat{\psi}(k, t) e^{i k \xi} dk$, we obtain
\begin{equation}
\mean{\xi}=  i \!\int  \mean{\hat{\psi}(k, t), \partial_k \hat{\psi}(k, t)}_{\sigma_z} dk,
\end{equation}
where we have used the identities $2\pi \delta'(k_2-k_1)=\int i x e^{i(k_2-k_1)x} dx$ and 
$\int \delta'(x-a) f(x) dx=-f'(a)$. Take the 
component $\int \psi_1^* \xi \psi_1 d\xi$ as an example, 
\begin{align}
\int \psi_1^* \xi \psi_1 d\xi &= 
\frac{1}{2\pi} \int \hat{\psi}_1^*(k_1) \hat{\psi}_1(k_2) \xi e^{i(k_2-k_1)\xi} d\xi dk_1 dk_2 \nonumber  \\
&=\frac{1}{i} \int \hat{\psi}_1^*(k_1) \hat{\psi}_1(k_2) \delta'(k_2-k_1) dk_1 dk_2 \nonumber \\
&= i \int \hat{\psi}_1^*(k) \partial_k \hat{\psi}_1(k) dk.
\end{align}
We split $\hat{\psi}(k, t)$ into two parts, i.e., $\hat{\psi}(k, t) = c_p e^{-i \epsilon t/\hbar} +  c_n e^{i \epsilon t/\hbar}$ where $c_p=\hat{f}(k) u_+(k)$ and $c_n=\hat{g}(k) u_-(k)$ represent the wavefunctions that have positive and negative energies,
respectively.  
Furthermore, we arrive at
\begin{equation}
\mean{\xi}= \xi_0 +  v_0 t + Z(t)
\end{equation}
where $\xi_0$ is the initial position. Using $\mean{u_\pm, u_\pm}_{\sigma_z}=\mp\epsilon/d_z$, the average velocity can be calculated by
\begin{equation}
v_0 =- \frac{1}{2 \hbar J_d }\int   (\partial_k \epsilon) \epsilon [\hat{f}^*(k)\hat{f}(k)+ \hat{g}^*(k)\hat{g}(k)]dk.
\end{equation}
The third term $Z(t)$ is given by
\begin{equation}\label{eq_zb}
Z(t) = i \!\int\! [ \mean{c_p,\partial_k c_n}_{\sigma_z} e^{-2i \epsilon t/\hbar} + \mean{c_n, \partial_k c_p}_{\sigma_z} e^{2i \epsilon t/\hbar} ] dk.
\end{equation}
In general Eq.~(\ref{eq_zb}) shows an oscillation, which is the so-called Zitterbewegung (ZB).
From Eq.~(\ref{eq_zb}) we can deduce that the coexistence of positive and negative energy states
is a necessary condition to find the ZB.

For $\hat{f}(k)$ and $\hat{g}(k)$ spectrally narrow at around $k=k_0$ the frequency of the ZB is established
as $\omega_{zb}=2\epsilon(k_0)/\hbar$. Especially for the case $k_0 \sim 0$, we have $\omega_{zb} \simeq 2\sigma/\hbar$.
For quasi-lD antiferromagnetic compound CsMnI$_3$, the typical parameters~\cite{Ivanov1995} are $S=5/2$,
$J/\hbar \simeq 198$ GHz and $D/\hbar \simeq 1.07$ GHz. We can establish the frequency of ZB
is $f_{zb} \simeq 32.8$ GHz.

\begin{figure}[tbhp]
\begin{center}
\includegraphics[width=0.48\textwidth]{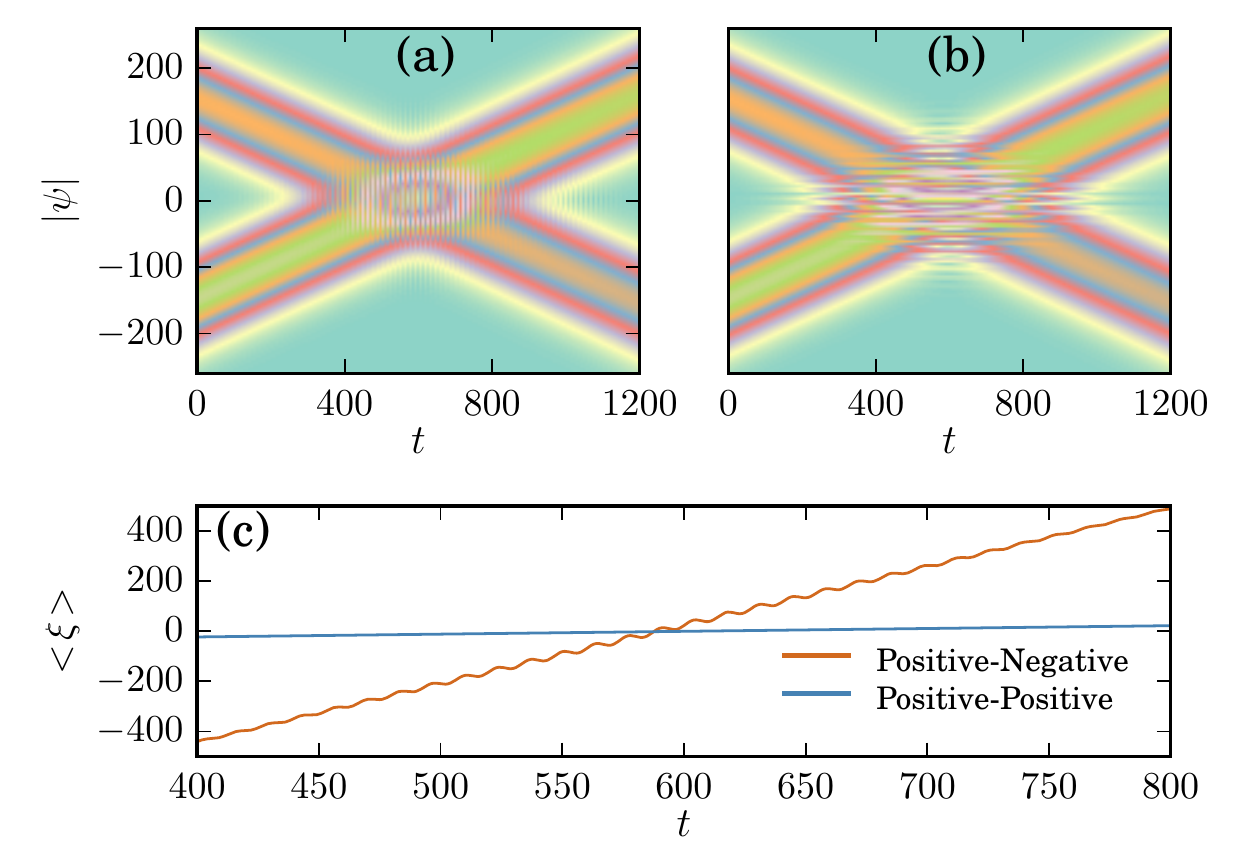}
\caption{(a) A positive wave packet and a negative packet that both have positive wave vector
move to different directions. The amplitude $|\psi|$ is used for plotting.
(b) Two positive wave packets with different wave vectors move toward each other too.
(c) The normalized average position $\mean{\xi}$ as a function of time for the two cases.}
\label{fig_zb}
\end{center}
\end{figure}

Fig.\ref{fig_zba}(b) shows the time evolution of the normalized expectation value of
position operator $\xi$, i.e., $N^{-1}  \mean{\xi}(t)$ as a function
of time where $N$ is the normalization factor. Clearly, in the long run
$\mean{\xi}$ follows a linear motion. However, a clear trembling motion is observed as well
in the initial stage ($t<100$).
This oscillating motion is the ZB. A discrete Fourier transform of $\mean{\xi}$ shows that the frequency is $0.095$,
which agrees with the predicted frequency $f_{zb}=\omega_{zb}/(2\pi) = 0.09$ very well.

In the previous example, the ZB decays gradually because the two packets are moving farther away
from each other in position space. In the following, we will prepare two individual Gaussian wave
packets and the initial profile is given by Eq.~(\ref{eq_sol}) such that $\psi_0(\xi)= \psi(\xi,0)$.
Here we choose $\hat{f}(k)= \hat{f}_1(k)$ and $\hat{g}(k)= \hat{f}_2(k)$ where
$\hat{f}_1(k) = A_1 e^{- \beta (k-k_1)^2} e^{-i \xi_1 k}$ and
$\hat{f}_2(k)=A_2 e^{- \beta (k-k_2)^2} e^{-i \xi_2 k}$ with $A_1=\sqrt{2\pi}/2$, $A_2=0.8 A_1$, $\beta =4000$,
$k_1=0.12$, $\xi_1=-300$, $k_2=0.12$ and $\xi_2=300$.
The packet located at $\xi_1=-300$ has a positive energy and the other has a negative energy with $\xi_2=300$.
Fig.~\ref{fig_zb}(a) depicts the evolution of them as a function of time,
apparently, two wave packets move towards each other and the interference occurs when they meet.
The corresponding normalized expectation value of position $\mean{\xi}$ is shown in Fig.~\ref{fig_zb}(c).
An oscillation motion (ZB) is found based on the linear motion. The ZB first increases then decreases
as the distance between the two packets changes.

As a comparison, we initialize the two wave packets with positive energies but one
of the wave vector $k$ is negative, i.e, $\hat{f}(k)=\hat{f}_1(k)+\hat{f}_2(k)$ and $\hat{g}(k)=0$ with $k_2=-0.12$.
In this case, ripples are generated but the pattern is different, as shown in Fig.~\ref{fig_zb}(b),
see the animation [II.gif] in the Supplemental Material~\cite{Supp}.
However, the normalized expectation value of position $\mean{\xi}(t)$ is a linear function of time.
As expected by Eq.~(\ref{eq_zb}), there is no ZB in this scenario since $\hat{g}(k)=0$.

We note that a Dirac-like magnon spectrum on a two-dimensional honeycomb lattice is reported 
recently~\cite{Fransson2016, Banerjee2016}, the corresponding Zitterbewegung should also occur on that systems.

\subsection{Klein paradox}
We now turn to the possible analog of the Klein paradox. Starting from the Dirac equation,
Klein calculated the reflection ($R_s$) and transmission ($T_s$) coefficients of an electron
incident on a step potential, and he found that the reflection coefficient $R_s>1$ for a larger potential,
this phenomenon is called Klein paradox~\cite{Hansen1981, Dombey1999, Calogeracos1999}.
This paradox can be resolved, that is, the relation $R_s+T_s=1$ still holds if one selects
the correct sign of momentum because for positrons the positive group velocity is associated
with a negative momentum. The nonzero transmission coefficient for strong potential, often referred
as Klein tunneling, shows that an incoming electron can penetrate through a high and wide potential
barrier without the exponential decay~\cite{Dombey1999, Calogeracos1999, Krekora2004, Katsnelson2006, Longhi2010a}.
The explanation of this relativistic effect regarding electron-positron production is that the positron
states can be excited inside the high barrier~\cite{Calogeracos1999, Katsnelson2006}.

\begin{figure}[tbhp]
\begin{center}
\includegraphics[width=0.48\textwidth]{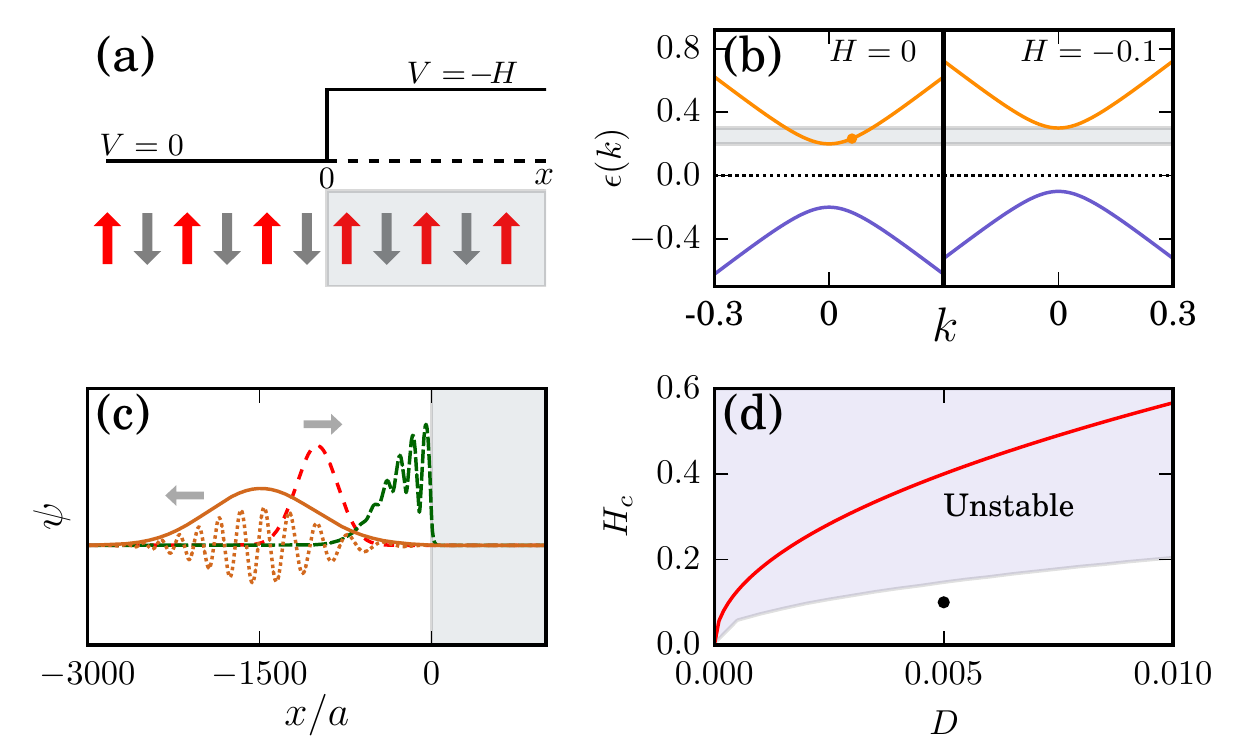}
\caption{(a) Schematic of the antiferromagnetic spin chain in a spatial Zeeman field,
which induces a potential $V=-H$ for $x>0$. (b) Dispersion relations for $H=-0.1$ with
parameter $D=0.005J$, the potential ($V=0.1$) is impenetrable for $k=0.06$ so the incident
wave packet will be totally reflected.
(c) Snapshots of the incident wave packet scatter with the Klein step.
Initially, a wave packet that has positive energy moves toward right (red dashed line),
then it is totally reflected due to the potential ($V=0.1$) and moves toward left.
(d) The critical field $H_c=2\sigma$ of the Klein tunneling as a function of $D$, which
is located in the unstable region.  The black point $(0.005, 0.1)$ is in the stable region.
}
\label{fig_klein}
\end{center}
\end{figure}

In the 1d antiferromagnetic system, we can set up a tunneling experiment by constructing a step potential
using spatial-varying external fields. Fig.\ref{fig_klein}(a) shows the schematic of the antiferromagnetic
spin chain in an external fields for $x>0$, and the corresponding potential $V=-H$. The dispersion relations
in the presence of external field are shown in Fig.\ref{fig_klein}(b). As we can see,
in the region $|H|<2\sigma$ the positive and negative branches do not overlap,
therefore, spin waves with the energy $\sigma<\epsilon(k)<\sigma-H$ will be totally reflected.
In Fig.\ref{fig_klein}(b) the gray bar plots the energy gap $[\sigma,\sigma-H]$.
As the potential increases, in the region $V>H_c=2\sigma$ the negative and positive energy branches
overlap, which leads to the so-called Klein tunneling.

The numerical simulation results are shown in Fig.\ref{fig_klein}(c). We initialize a wave packet
with positive energy and $k_0=0.06$, as shown in Fig.\ref{fig_klein}(c) with a red dashed line.
The wave packet moves toward the right, it gets totally reflected when meets the external field barrier ($V=0.1$).
The Klein tunneling requires a large external field ($|H|>H_c$), it is necessary to check whether
the system is still in its ground state. Large external fields will push the system out of the stable state.
For instance, a large field could induce the spin-flop phase in which each spin almost antiparallel
with its neighbors and approximately perpendicular to the external field~\cite{Rezende2016}.
The unstable region displayed in Fig.\ref{fig_klein}(d) is obtained numerically by comparing several possible states.
The critical field $H_c$ as a function of $D$ is plotted in Fig.\ref{fig_klein}(d), and we find
that $H_c$ is located in the unstable region. Therefore, we conclude that the Klein tunneling
is not accessible in this antiferromagnetic system.

\section{Summary}
In summary, we have studied the magnon excitations in a classical antiferromagnetic spin chain
in the presence of easy-axis anisotropy. We obtained a Dirac-like equation by linearizing the LLG equation.
Based on the Dirac-like equation we show analytically that the classical analog of relativistic ZB can occur
in the classical Heisenberg antiferromagnetic spin chain. We describe it as the magnonic analog of
relativistic Zitterbewegung and have demonstrated it by solving the LLG equation numerically.
We also discussed the magnonic analog of Klein paradox and found that the Klein tunneling region
is inaccessible because a large external field will break the antiferromagnetic ground state.

We acknowledge the financial support from National Natural Science Foundation of China (Grant No. 11604169) and
EPSRC under Centre for Doctoral Training Grant EP/L015382/1. This work is sponsored by an open fund (No. xkzwl1614) 
and K.C.Wong Magna Fund in Ningbo University.

\bibliographystyle{apsrev4-1}
%

\end{document}